\documentclass[sigconf,nonacm]{acmart}

\AtBeginDocument{%
  }

\setcopyright{acmlicensed}
\copyrightyear{2018}
\acmYear{2018}
\acmDOI{XXXXXXX.XXXXXXX}

\acmConference[Conference acronym 'XX]{Make sure to enter the correct
  conference title from your rights confirmation email}{June 03--05,
  2018}{Woodstock, NY}

\acmISBN{978-1-4503-XXXX-X/2018/06}

\acmSubmissionID{2655}

\providecommand{\shownotes}{0} 

\ifnum\shownotes=1
  \newcommand{\note}[1]{\par\smallskip{\color{blue!70!black}\small[\,#1\,]}\par\smallskip}
  
\else
  \newcommand{\note}[1]{}
  
\fi

\newcommand{\toolname}{\textsc{Vibe-GUIDE}}

\usepackage{colortbl}
\usepackage{placeins}
\definecolor{primaryrolebg}{HTML}{EAF1F6}
\definecolor{secondaryrolebg}{HTML}{FAF0E2}
\definecolor{exploratoryrolebg}{HTML}{F2F3F4}
\definecolor{graphcondition}{HTML}{173F5F}
\definecolor{chatcondition}{HTML}{C4862C}
\newcommand{\primaryrole}{\cellcolor{primaryrolebg}\textbf{Primary}}
\newcommand{\secondaryrole}{\cellcolor{secondaryrolebg}Secondary}
\newcommand{\exploratoryrole}{\cellcolor{exploratoryrolebg}Exploratory}
\PassOptionsToPackage{table}{xcolor}

\usepackage{enumitem}
\usepackage[english]{babel}
\addto\extrasenglish{%
}

\begin{document}

\title{Vibe-GUIDE: A Graph-based User Interface in IDEs for Oversight in Vibe Coding}

\definecolor{hilite}{rgb}{0.99,0.92,0.85}

\author{Chifang Chou}
\email{cfchou@ucdavis.edu}
\affiliation{%
  \institution{University of California, Davis}
  \city{Davis}
  \state{California}
  \country{USA}}

\author{Sam Yu-Te Lee}
\email{ytlee@ucdavis.edu}
\affiliation{%
  \institution{University of California, Davis}
  \city{Davis}
  \state{California}
  \country{USA}}

\author{Rudrajit Choudhuri}
\email{choudhru@oregonstate.edu}
\affiliation{%
  \institution{Oregon State University}
  \city{Corvallis}
  \state{Oregon}
  \country{USA}}

\author{Kwan-Liu Ma}
\email{klma@ucdavis.edu}
\affiliation{%
  \institution{University of California, Davis}
  \city{Davis}
  \state{California}
  \country{USA}}

\renewcommand{\shortauthors}{Chou et al.}

\begin{abstract}
In agentic coding, developers shift from implementing changes themselves to specifying intent, evaluating the agent’s work, and making approval decisions. However, delegating implementation can introduce cognitive debt that erodes project comprehension over time, constraining developers’ ability to provide oversight. In this work, we investigate the role of persistent shared representations in supporting project comprehension and oversight of coding agents. We present Vibe-GUIDE, an agentic coding interface built around a structural, live, manipulable, and adaptive graph representation organized by functional modules. We evaluated our interface in a randomized between-subjects study comparing how 16 developers completed three cumulative coding tasks using our interface or a Chat-only baseline. We found that Vibe-GUIDE can support project comprehension and sustained task performance while keeping developers cognitively involved in oversight. These findings show how persistent shared representations can complement natural-language interaction and help developers maintain the understanding needed to oversee agent-generated changes as projects evolve.
\end{abstract}



\maketitle

\section{Introduction}

AI coding agents can inspect a project, edit files, and run development tools while carrying out multi-step requests~\cite{watanabe2026useagenticcodingempirical}. In this form of \emph{agentic coding}, developers shift from implementing each requirement themselves toward specifying intent, evaluating the agent's work, and deciding whether to accept, redirect, or revise it~\cite{waseem2025vibecodingpracticeflow,geng2026interactionofvibecoding}. This is also described as \emph{vibe coding}, particularly when developers rely on natural-language instructions and accept generated code without closely inspecting it~\cite{sarkar2025vibecodingprogrammingconversation}.

This shift in how developers work creates a challenge for \emph{project comprehension}: developers' working understanding of how and why a system operates and how its parts relate. As agents take over implementation, developers can accept changes without tracing how they work or fit into the project, reducing opportunities to build the understanding ordinarily developed through programming~\cite{naur1985programmingastheorybuilding}. Over time, this gap can accumulate into cognitive debt: the erosion of the understanding needed to explain, evaluate, and maintain the project~\cite{storey2026technicaldebtcognitiveintent}.

Major AI governance frameworks call for human oversight in agentic coding~\cite{tabassi2023airmf,oecd2024aiprinciples}, and developers themselves prefer to retain oversight of AI-produced work~\cite{choudhuri2026youshallnotpass}. Yet to decide whether to accept, redirect, or correct an agent, developers need enough understanding of the project to judge where its changes fit and what else they may affect.
The growing project comprehension gap can thus constrain their ability to provide oversight.

To support project comprehension and oversight, we explore a shared representation design~\cite{heer2019agencyautomation} through which developers and agents can act throughout agentic coding sessions. Drawing on prior research, we formulated four design requirements: the representation should be (1) \emph{structural}, organizing the project at the problem-solving level~\cite{brooks1983theoryofcomprehensionofcomputerprograms,soloway1984empiricalstudiesofprogrammingknowledge}; (2) \emph{live}, remaining synchronized with agent-generated changes~\cite{xie2024waitGPT,horowitz2023liverichcomposablequalities}; (3) \emph{manipulable}, allowing developers to steer the agent through represented modules and relationships~\cite{shneiderman1983directmanipulation,zhang2025neurosync}; and (4) \emph{adaptive}, revealing additional detail about the developer’s current focus~\cite{shu2023sensecape,lee2025HINT}. Instantiating these requirements, we developed \toolname{}, an agentic coding interface based on a graph representation of the project. The graph provides a project-level abstraction organized around functional modules---components defined by what the software does---and their relationships, rather than mirroring the codebase’s directory structure. It remains synchronized with the underlying codebase and provides an additional surface for inspecting and directing the agent’s work.

To evaluate this design, we conducted a randomized, between-subjects study with 16 participants. All participants completed three cumulative project-modification tasks with the same coding agent, with eight assigned to the Graph condition and eight to the Chat-only condition, and then completed a project-comprehension assessment. We ask:

\begin{itemize}
  \item[\textbf{RQ1}] How does using \toolname{} during agentic coding affect developers' project comprehension, task performance, cognitive load, and trust in the coding agent?
  \item[\textbf{RQ2}] How do developers use \toolname{}'s graphical representation to oversee agent-generated changes during their creation and subsequent review?
\end{itemize}

The user study finds that both groups completed every task and achieved similar primary project-comprehension scores. Graph participants demonstrated more relationship-oriented comprehension and maintained a steadier pace across the cumulative tasks, alongside higher mental demand, lower effort, similar frustration, and lower trust. Participants used the Graph primarily to inspect project structure and agent-generated changes, while chat remained their main channel for directing the agent and conventional development resources supported verification. We discuss what these findings suggest about the four design requirements, how graph-based shared representations can complement natural-language interaction and support human oversight, and how the design might extend to other software-engineering activities. Future work should isolate the contributions of the four design requirements and evaluate their longer-term use across varied projects and tasks.

This work makes three contributions:
\begin{itemize}
  \item four literature-grounded design requirements for shared representations that support project comprehension and human oversight during agentic coding;
  \item \toolname{}, an agentic coding interface that instantiates these requirements through a graph-based shared representation; and
  \item findings from a randomized exploratory evaluation, characterizing how developers used the representation and the corresponding design trade-offs.
\end{itemize}

\section{Related Work}

This work connects three lines of research: agentic coding and developers' understanding, human oversight of agents, and interactive representations for sensemaking and steering. Together, they frame our investigation of a shared representation through which developers can inspect and modify an evolving codebase while working with a coding agent.

\subsection{Agentic Coding and Cognitive Debt}
We use \emph{agentic coding} to refer to software development in which a developer delegates multi-step work to an AI agent that can inspect a codebase, edit files, and run development tools~\cite{watanabe2026useagenticcodingempirical}. Compared with conventional development, developers shift from implementing requirements to specifying intent, evaluating the agent's work, and deciding whether to accept, redirect, or revise it~\cite{waseem2025vibecodingpracticeflow,geng2026interactionofvibecoding}. This role emphasizes context management, rapid evaluation, and coordination with the agent~\cite{sarkar2025vibecodingprogrammingconversation}.

Recent studies show that agentic coding can provide short-term efficiency gains while increasing verification effort and longer-term risks. Pimenova et al.~\cite{pimenova2026goodvibrationsqualitativestudy} identify code review, reliability, and debugging as recurring challenges, while Fan et al.~\cite{fan2026whenhelphurts} found that verification load contributed to stress and fatigue across repeated tasks. These costs can reside in the artifact as technical debt~\cite{cunningham1992techdebtorigin,he2026cursor,liu2026debtaiboomlargescale} or in developers' diminishing understanding of it.

Our work focuses on \emph{project comprehension}: developers' working understanding of how and why a system operates. Naur~\cite{naur1985programmingastheorybuilding} describes programming knowledge as a theory built through work on a system; agentic coding can weaken this process when developers accept changes without reconstructing how they fit into the codebase. Storey~\cite{storey2026technicaldebtcognitiveintent} calls the accumulation of lost understanding \emph{cognitive debt}. Related formulations include \emph{comprehension debt}, the gap between what a team knows and must know to maintain a codebase~\cite{ahmad2026comprehensiondebtgenaiassistedsoftware}, and \emph{epistemic debt}, which emphasizes developers' ability to explain or repair AI-assisted programs~\cite{sankaranarayanan2026mitigatingepistemicdebtgenerative}. We treat these as related manifestations of cognitive debt and focus on project comprehension. Whereas this literature names and characterizes losses of understanding, we examine whether an interface can help developers preserve comprehension as agent-generated changes accumulate.

\subsection{Human Oversight and Epistemic Access}

Human oversight involves monitoring an AI system and intervening to mitigate task-specific risks~\cite{gaube2026keepinganeye}. In agentic coding, oversight depends on project comprehension: developers need sufficient understanding of the project to determine whether the agent's actions or outputs warrant intervention. Dhanorkar et al.~\cite{dhanorkar2026humanoversight} identify a priori control, co-planning, real-time monitoring, and post hoc review, yet their participants still found generated code difficult to review and often relied on heuristics such as test results. These practices assess behavior and outputs without necessarily revealing how a change fits into the surrounding codebase.

Interfaces can support oversight by providing \emph{epistemic access}: task-relevant visibility into the agent's activity and effects~\cite{sterz2024effectivenessinhumanoversight}. Such visibility must be selective. Passi~\cite{passi2025oversight} argues that showing all activity is infeasible and that useful visibility depends on the user, task, and context; Mitchell et al.~\cite{mitchell2026aiagentspush} similarly warn that agent systems can erode the cognitive capacities on which oversight depends. In code completion, Vasconcelos et al.~\cite{vasconcelos2025generation} found that highlighting tokens likely to require editing supported faster, more targeted edits, whereas highlighting based only on generation probability did not. Oversight therefore depends less on exposing more information than on making relevant information accessible at the right time. Rather than supporting oversight by exposing more agent activity, our work makes the structural effects of agent actions visible when developers evaluate or redirect them.

\subsection{Interactive Representations for Sensemaking and Steering}

Interactive external representations can provide epistemic access by making task-relevant relationships perceptually available rather than requiring users to reconstruct them from a linear information stream~\cite{1987whyadiagramworths,1996externalcognition}. LLM interfaces apply this principle through spatial and graph-based workspaces that preserve context across navigation and levels of detail~\cite{shu2023sensecape,ma2025gardenofpapers,lee2025HINT}.

For project comprehension, representations must connect developers' domain goals to program structures~\cite{brooks1983theoryofcomprehensionofcomputerprograms,soloway1984empiricalstudiesofprogrammingknowledge}. Ivie~\cite{yan2024ivie} connects explanations at different levels of detail to generated code, while CodeVoyager~\cite{kim2026codevoyager} combines natural-language queries with interactive call and control-flow graphs. Both support inspection and navigation but do not maintain a shared representation as an agent modifies the project.

Representations can also support intervention. VizCopilot~\cite{lee2025vizcopilot} visualizes and lets users select chatbot context. DirectGPT~\cite{masson2024directgpt} and WaitGPT~\cite{xie2024waitGPT} translate manipulations of generated objects or operations into prompts, while NeuroSync~\cite{zhang2025neurosync} and Just-in-Time Objectives~\cite{lam2026justintimeobjectives} expose inferred tasks or objectives for revision. These systems show how direct manipulation can complement language by making elements of an AI interaction inspectable and editable.

Prior systems thus support complementary parts of the problem: preserving context, connecting explanations to code structure, or externalizing objects for steering. They leave open how one persistent representation can support both comprehension and steering as an agent changes a codebase. Our design combines these functions in a live, adaptive, directly manipulable representation synchronized with the evolving project.

\section{Design Requirements}

Drawing on prior work on cognitive debt, human oversight, and interactive representations, we formulated four requirements for an interface that helps developers maintain project comprehension while working with a coding agent. Together, these requirements define a representation that is \emph{structural} (organized around what the code does; DR1), \emph{live} (synchronized with codebase changes; DR2), \emph{manipulable} (editable as a means of steering the agent; DR3), and \emph{adaptive} (able to reveal detail based on the developer's current focus; DR4).

\begin{description}[style=nextline, leftmargin=1.5em, labelindent=0pt]
    \item[DR1: Provide a structural view at the problem-solving level:]
    The representation should organize the project around what the software does and how its major components relate. We call this the \emph{problem-solving level}: it connects domain goals to program structures rather than mirroring source files or code text~\cite{brooks1983theoryofcomprehensionofcomputerprograms,soloway1984empiricalstudiesofprogrammingknowledge}. Exposing this mapping can help developers interpret generated changes within the broader codebase and maintain the understanding that agentic coding may otherwise erode~\cite{storey2026technicaldebtcognitiveintent}.

    \item[DR2: Keep the representation live and synchronized:]
    The representation should be \emph{live}: it should update as the agent modifies the codebase and make recent changes visible. A static snapshot can diverge from the evolving artifact, requiring developers to reconstruct what changed from code or conversation history. Maintaining synchronization can preserve context across successive iterations and support review of the agent's work~\cite{xie2024waitGPT,horowitz2023liverichcomposablequalities}.

    \item[DR3: Make the representation directly manipulable:]
    A directly manipulable representation should let developers express intent by adding, modifying, removing, or connecting represented elements, with the agent carrying those requests into the codebase. Direct manipulation complements natural-language prompting by allowing users to act on externalized objects and relationships~\cite{shneiderman1983directmanipulation,masson2024directgpt,xie2024waitGPT,zhang2025neurosync}. This capability turns the representation from a read-only report into an interface for steering the agent.

    \item[DR4: Adapt the representation to the developer's current focus:]
    An adaptive representation should preserve a project-level overview while revealing additional detail about the components relevant to the developer's current task. Interfaces that organize information across levels of abstraction can support focused exploration without presenting all available detail at once~\cite{shu2023sensecape,lee2025HINT,zhang2025neurosync}. Adapting the visible detail keeps the structural overview in DR1 usable as the codebase grows and the interaction evolves.
\end{description}

Together, DR1 and DR4 determine how the representation supports project comprehension, while DR2 and DR3 make it bidirectional: agent-made code changes appear in the representation, and developer-made changes to the representation are carried back into the codebase by the agent. The result is a shared working surface that conveys information and intent in both directions rather than static documentation. \autoref{sec:system-design} describes how \toolname{} realizes these requirements.

\section{\toolname{}: System Design}\label{sec:system-design}

This section describes how \toolname{}\footnote{https://anonymous.4open.science/r/Vibe-GUIDE} instantiates the four design requirements through an Overview Graph that complements the agent conversation.

\begin{figure*}[t]
  \centering
  \includegraphics[width=\textwidth]{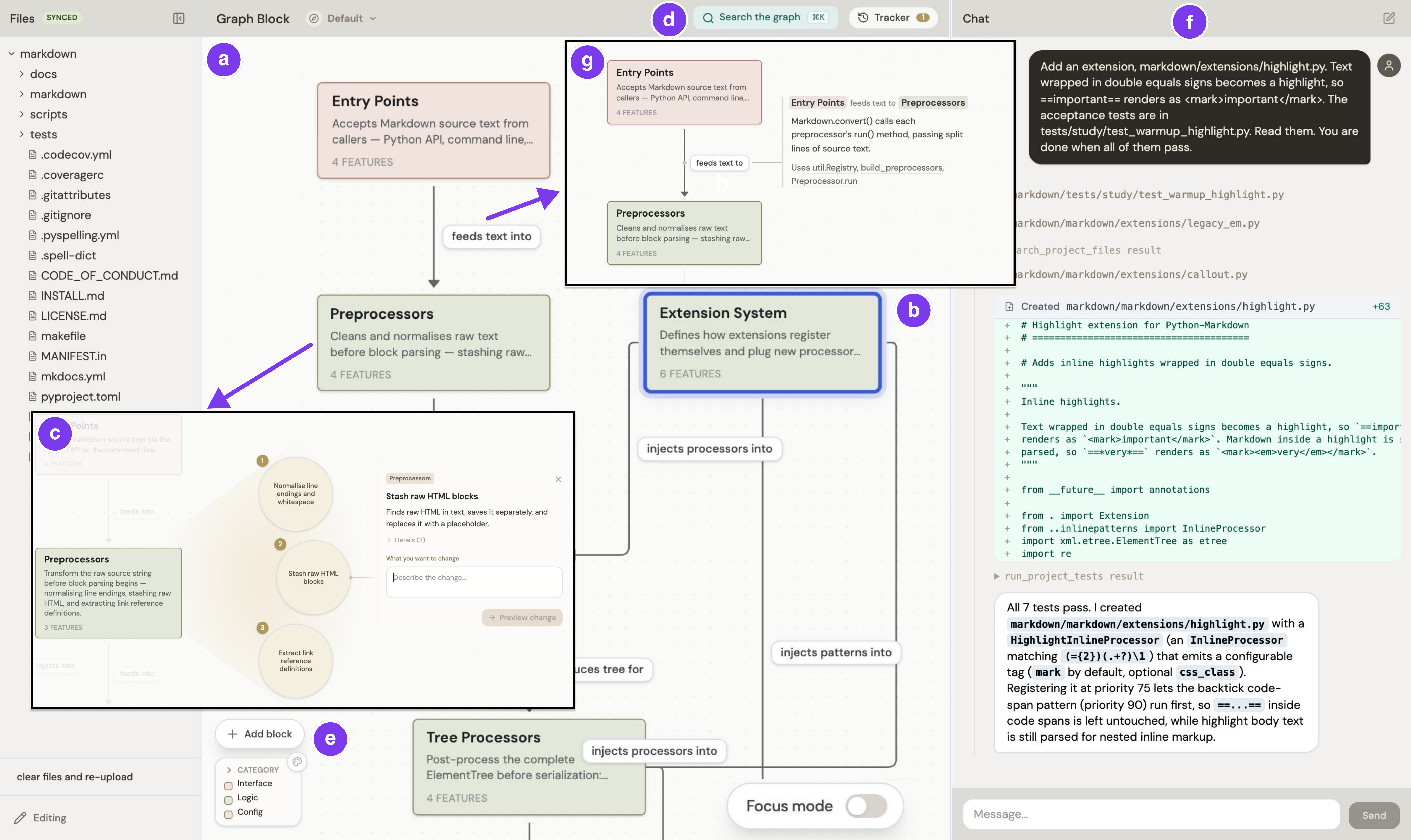}
  \caption{\toolname{} interface with a project open. (a) The Overview Graph is a node-link view of the project. (b) A solid
blue ring marks a module the agent has just changed, and it stays until the
developer's next action. (c) Clicking a module opens its feature bubbles, and opening a bubble gives more details and a box where the developer describes the change they want. (d) The search toolbar filters modules by name or takes a question in natural language, and the tracker beside it lists what modules or arrows the agent has changed. (e) A module can also be added by hand without using the chat. (f) The Chat Panel shows the agent's tool calls and replies. (g) Clicking an arrow explains in one sentence how that connection works in the code.}
\Description{The interface has three columns, a file tree on the left, the Overview Graph in the middle, and the Chat Panel on the right. The file tree lists the folders and files of an open project. The Graph shows four modules in a single downward chain, Entry Points, then Preprocessors, then Extension System, then Tree Processors. Each module is a rounded card carrying a name, a one-sentence description, and a count of features. The arrows between them are labelled in plain words, reading feeds text into, injects processors into, and injects patterns into. Two enlarged insets are laid over the canvas. One shows a module fanned open into four numbered bubbles, with one bubble expanded into a card that holds a short explanation and a text box reading What do you want to change. The other shows an arrow explanation naming the two modules it joins and the calls behind it. A toolbar above the canvas holds a search box and a tracker. A button reading Add block sits below the canvas, and a Focus mode switch sits at its lower right. The Chat Panel shows a typed instruction asking for a highlight extension, then the files the agent read, then the contents of the file it created, then a summary saying all seven tests pass.}
  \label{fig:overview}
\end{figure*}

\subsection{System Overview}\label{sec:tech}

\toolname{} combines a file tree, the \emph{Overview Graph}, and a \emph{Chat Panel} (\autoref{fig:overview}). The Chat Panel provides a conventional interface to the coding agent, while the Graph maintains a project-level representation beside the conversation. The two are not independent views: developers can initiate changes through either one, and the coding agent's activity returns to the Graph. A backend service runs the coding agent, while a separate graph-generation model constructs and updates the Graph. This bidirectional loop makes the Graph a shared surface for project comprehension and oversight.

\subsection{Constructing the Overview Graph}

\paragraph{Representation and visual encoding.}
The central design choice is to organize the Graph by project functionality rather than source-code structure. A \emph{module} represents a functional part of the project, \emph{features} describe behaviors within it, and \emph{relationships} show how modules work together. This problem-solving-level organization connects what the project does with how its parts interact without reproducing the file tree (DR1).

Modules appear as cards with plain-language labels, short descriptions, and colors indicating their roles (\autoref{fig:overview}-a). Numbered bubbles reveal their features, while labeled arrows express relationships using verbs such as ``drives'' or ``feeds into.'' The top-down layout gives the Graph a consistent reading direction, and selecting a feature or relationship reveals more detail (\autoref{fig:overview}-c,g).

\paragraph{Graph generation.}
To construct the Graph, the server sends the project files to Claude Sonnet 4.6 with a fixed prompt. The prompt asks the model to produce structured descriptions of modules, features, relationships, and their mappings to files and functions. These mappings connect the functional representation to the underlying code and later allow \toolname{} to associate agent activity with Graph elements. The server streams generated elements to the interface, which renders them with React Flow~\cite{reactflow} and lays them out with dagre~\cite{dagre}.

\paragraph{Designer-defined structure and model-inferred content.}
The interface fixes the Graph's node-link form, layout, role encoding, and progressive disclosure; the graph-generation model infers its modules, features, relationships, and mappings to files and functions. The prompt also asks the model to use functional labels rather than names copied from the file tree.

A project-level overview creates a tension between structural coverage and visual complexity. The prompt therefore targets no more than eight modules, while feature bubbles (\autoref{fig:overview}-c) and Focus Mode (\autoref{fig:scenario2}-e) reveal finer detail on demand. This progressive disclosure preserves the Overview Graph as a stable point of orientation while providing additional detail when needed (DR4).

\subsection{Interaction and Coordination}\label{sec:coord}

The Graph becomes a shared representation through two movements. Developers use its structure to inspect the project and direct the coding agent; the interface projects the agent's activity and resulting changes back onto that same structure.

\subsubsection{Inspecting Project Structure}

Selecting a module reveals its features, while selecting a feature or relationship provides a more detailed explanation (\autoref{fig:overview}-c,g). The search toolbar either filters visible elements by name or converts a natural-language question into an ordered reading path through relevant modules and relationships (\autoref{fig:overview}-d and \autoref{fig:scenario1}-a). Together, these interactions let developers move from a behavioral question to the corresponding project structure without first knowing the source-code vocabulary (DR1).

Focus Mode adapts this structure to the current conversation. It identifies the modules relevant to a recent request and generates a finer-grained graph around them (\autoref{fig:scenario2}-e). Developers can inspect these \emph{detail modules} or promote selected ones to the Overview Graph (\autoref{fig:scenario2}-f), preserving details that become important to their work (DR4).

\subsubsection{Steering the Coding Agent}

The Graph also provides an alternative entry point for directing the coding agent. Developers can act on a module, feature, or relationship and either describe the intended change or ask the coding agent for suggestions (\autoref{fig:overview}-c,e and \autoref{fig:scenario1}-b). The selected element supplies the structural context, while the developer's description or chosen suggestion supplies the requested behavior (DR3).

Confirmed requests are sent to the coding agent as \emph{Graph edits} and recorded in the Chat Panel alongside typed prompts, tool calls, and agent responses (\autoref{fig:scenario1}-c). The conversation therefore preserves how each structural action was translated into a coding request.

\subsubsection{Following Agent Changes}

\toolname{} distinguishes where the coding agent is currently working from what it ultimately changed. Modules associated with files being read or edited blink during execution. When the turn ends, elements associated with completed modifications remain highlighted until the developer begins another request (\autoref{fig:overview}-b and \autoref{fig:scenario2}-d).

The change tracker records added, modified, and removed Graph elements and lets developers return to each one (\autoref{fig:overview}-d). After code changes, the graph-generation model receives the updated codebase and the previous Graph, allowing it to incorporate new behavior while preserving stable project structure where possible. These mechanisms make agent activity visible at the same functional level developers use to inspect and steer the project (DR2).

\autoref{sec:usecase} illustrates how these inspection, steering, and change-awareness mechanisms work together in a single agentic coding session.

\section{Illustrative Use Case}
\label{sec:usecase}

\begin{figure*}[t]
  \centering
  \includegraphics[width=\textwidth]{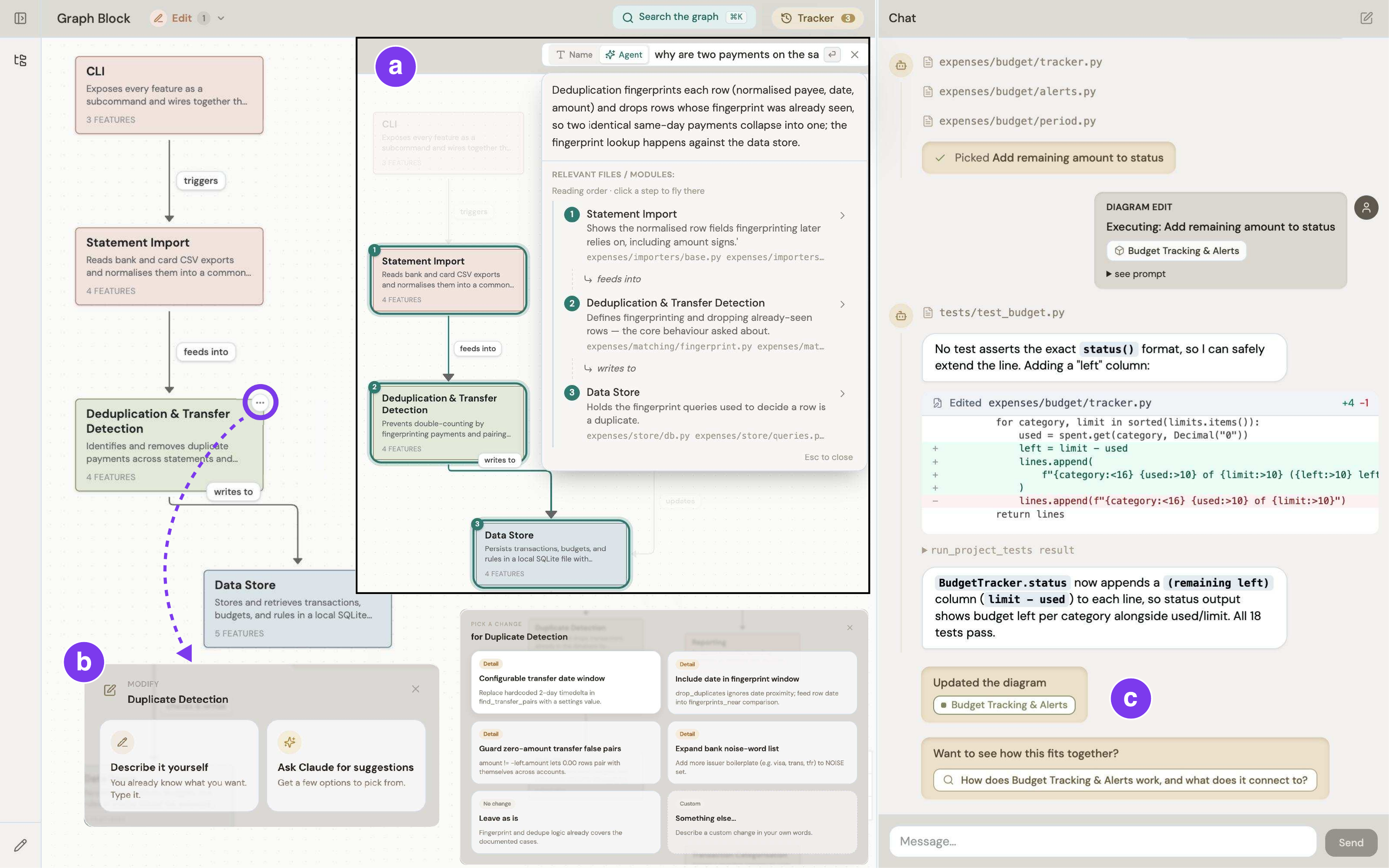}
  \caption{Ana's session in one view. (a) She searches the Graph in her own words, and the answer names the module responsible and numbers a three-step reading path across the canvas. (b) The actions button on a block (circled) opens two ways, describing the change herself or asking for suggestions, which show as cards on the canvas. (c) The modification requests are also recorded in the chat panel.}
  \Description{One screen with the Overview Graph on the left and the Chat Panel on the right. A search answer and a change dialog are laid over the Graph, which shows a personal finance project. Its modules are Command Line Interface, Statement Import, Deduplication and Transfer Detection, Data Store, Budget Tracking and Alerts, and Reporting and Export, joined by arrows labelled triggers, feeds into, and writes to. The search panel over the canvas answers a typed question about duplicate payments in prose, then lists three modules as numbered steps, Statement Import first, Duplication and Transfer Detection second, and Data Store third, with the matching modules numbered on the canvas behind it. Below the canvas a small dialog offers two choices, Describe it yourself and Ask Claude for suggestions, and beside it six suggestion cards appear, five naming a specific change and one reading Something else. In the Chat Panel, an entry marked Diagram Edit names the change being carried out and offers to reveal the prompt that was sent. Under it are the edited lines of one file, a summary saying all eighteen tests pass, a chip reading Updated the diagram followed by the name of the changed module, and a follow-up question offering to explain how that module connects to the rest.}
  \label{fig:scenario1}
\end{figure*}

\begin{figure*}[t]
  \centering
  \includegraphics[width=\textwidth]{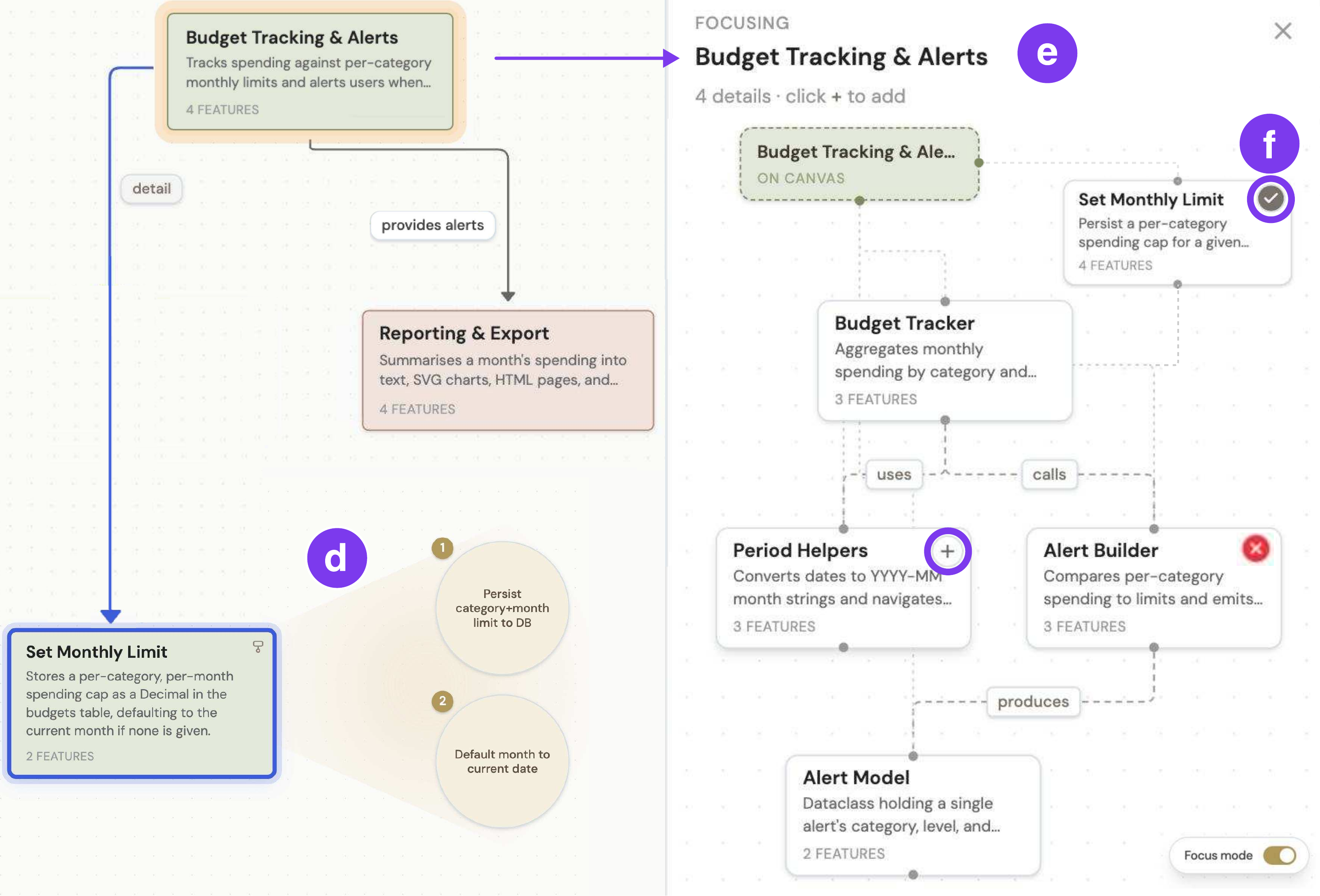}
  \caption{How Ana looks closer. (d) When the modification finishes, the canvas highlights what changed, here shows the block she added back to the overview. Opening a module fans it out into its numbered detail bubbles. (e) Focus mode narrows the canvas to the modules in play and grows the finer detail modules inside them. (f) Each of the modules can be added back to the Overview Graph.}
  \Description{A split view with the Overview Graph on the left and the Focus mode side panel on the right. On the left, the module named Budget Tracking and Alerts is drawn with a highlight to mark it as newly changed. An arrow labelled provides alerts runs from it to Reporting and Export, and a connector labelled detail runs down to a module named Set Monthly Limit, which is fanned open into two numbered bubbles. On the right, the side panel is headed with the name of the module in focus and a count of four details. Inside it, six smaller cards are drawn as their own graph. The card for the module in focus is marked as already on the canvas, and the others are Set Monthly Limit, Budget Tracker, Period Helpers, Alert Builder, and Alert Model. Dashed connectors between them are labelled uses, calls, and produces. Each detail card carries a control at its corner for adding it to the Overview Graph or dismissing it. A Focus mode switch at the lower right is turned on.}
  \label{fig:scenario2}
\end{figure*}

In this hypothetical scenario, we follow Ana, a developer taking over a small expense-tracking project she has never opened, to illustrate how \toolname{} can be used to find, understand, and change project behavior. Ana has two tasks. First, she must fix a bug that merges two separate payments made on the same day for the same amount. Second, she must improve the project's budgeting functionality without being told what change to make. She begins by uploading the project into \toolname{}, which generates the Overview Graph shown in~\autoref{fig:scenario1}.

Ana begins with the specific bug but does not know which part of the project is responsible. She therefore asks the search toolbar, ``Why are two payments on the same day for the same amount treated as one?'' The response identifies \emph{Duplication \& Transfer Detection} and lays out a three-step reading path across the Graph (\autoref{fig:scenario1}-a). This path is intended to help her locate the relevant project behavior without first knowing the names of its files or features.

Ana opens the identified module, which expands to show its four features (\autoref{fig:scenario2}-d). Selecting the relevant feature opens a card explaining why the payments are merged at the problem-solving level (DR1). From this information, Ana decides how to address the bug. She opens the module's actions and describes the fix she wants (DR3). The coding agent carries out the change, while the request and resulting file operations are recorded in the Chat Panel (\autoref{fig:scenario1}-c).

Having resolved the specific bug, Ana turns to the open-ended task. She opens the actions on \emph{Budget Tracking \& Alerts} and asks the coding agent for suggestions (\autoref{fig:scenario1}-b). Six cards appear on the canvas: five suggestions from the coding agent, including one that proposes making no change, and a sixth that lets Ana describe another option in her own words. She selects one of the suggestions. The affected module begins blinking while the coding agent works and remains highlighted when the modification is complete (\autoref{fig:scenario2}-d).

After the change, Ana turns on Focus Mode to inspect the modified module in greater detail (\autoref{fig:scenario2}-e). She promotes the \emph{Set Monthly Limit} detail module to the Overview Graph, where it remains after she closes Focus Mode (\autoref{fig:scenario2}-f). This lets her preserve a finer-grained part of the project that has become relevant to her work (DR4).

Together, the scenario shows how Ana moves from a behavioral question to a structurally situated code change, then preserves relevant detail for subsequent work. The following study compares this workflow with the same agentic coding interface without the Graph (\autoref{sec:study} and~\autoref{sec:results}).

\section{Comparative Evaluation}
\label{sec:study}

We conducted a randomized between-subjects evaluation comparing a \emph{Graph} group, which used the full \toolname{} interface, with a \emph{Chat-only} group, which used the same interface without the Graph panel.

The evaluation addressed two research questions:

\begin{description}[style=sameline, leftmargin=3.5em, labelwidth=3em, labelsep=0.5em, labelindent=0pt, align=left]
  \item[\textbf{RQ1}] How does using \toolname{} during agentic coding affect developers' project comprehension, task performance, cognitive load, and trust in the coding agent?
  \item[\textbf{RQ2}] How do developers use \toolname{}'s graphical representation to oversee agent-generated changes during their creation and subsequent review?
\end{description}

\subsection{Study Design}
\label{sec:study:design}

We randomly assigned participants by shuffling 8 assignments for each group. Participants were unaware of the other group. Both groups used the same application build, coding agent, tasks, and logging; the Graph group additionally received the interface functions described in \autoref{sec:system-design}. The comparison therefore isolates access to the bundled Graph interface.

\subsubsection{Participants}
\label{sec:study:participants}

We recruited participants through institutional mailing lists and the authors' professional networks. Eligibility required experience using AI coding agents and either formal programming training or a software-engineering job title. All 16 enrolled participants completed the study and were included in the evaluation, with 8 assigned to each group.

Participants ranged from 20 to 30 years old (median 25); 10 identified as men and 6 as women. They reported 2 to 10 years of programming experience (median 5.5 years). Fifteen used AI coding tools multiple times a day, 1 used them about once a week, and 13 had previously used Claude Code.

\subsubsection{Tasks}
\label{sec:study:tasks}

Participants completed an untimed warm-up followed by 3 cumulative study tasks in a fixed order (\autoref{tab:tasks}). The tasks used a study fork of Python-Markdown 3.10.3. Task~1 created a callout extension, Task~2 extended it into a panel extension, and Task~3 merged the panel and table-of-contents extensions. Each task began from the participant's current project state.

Each task provided a behavior-oriented brief and acceptance-test suite without naming relevant functions, classes, priorities, or data structures. Each task stage lasted 15 minutes. Participants who passed early used the remaining time to inspect task-related code, giving everyone equal exposure to each task. The task runner recorded acceptance-test results, elapsed time, and whether the complete suite passed.

\begin{table*}[t]
  \centering
  \caption{Evaluation tasks. Tasks~1--3 were completed in a fixed, cumulative sequence. Acceptance-test counts indicate task scope rather than comparable difficulty.}
  \label{tab:tasks}
  \begin{tabular}{@{}l p{0.58\textwidth} cc@{}}
    \toprule
    Task & Requested behavior & Acceptance tests & Time limit \\
    \midrule
    Warm-up & Render inline highlights written as \texttt{==text==} with a
              \texttt{<mark>} element & 7 & Untimed \\
    1 & Create typed and numbered callout blocks with anchors and 2-way cross-references & 31 & 15 min \\
    2 & Turn callouts into panels, adding back-references and numbering that restarts at headings & 45 & 15 min \\
    3 & Merge the panel and table-of-contents extensions and add heading references & 64 & 15 min \\
    \bottomrule
  \end{tabular}
  \Description{A four-column table listing the warm-up and the three timed tasks. The columns are the task name, the requested behaviour, the number of acceptance tests, and the time limit. The warm-up has seven acceptance tests and no time limit. Tasks one, two and three have thirty-one, forty-five and sixty-four acceptance tests, and each has a fifteen-minute limit. Each task extends the Markdown extension built in the task before it.}
\end{table*}

\subsection{Measures and Analysis}
\label{sec:study:measures}
\label{sec:study:analysis}

\autoref{tab:measures} summarizes the measures and their analytic roles. We did not select the sample size through an a priori precision analysis. Given our small exploratory sample and the ASA's caution that $p$-values neither convey effect magnitude nor justify conclusions on their own~\cite{wasserstein2016asa}, we follow prior HAI evaluation practice~\cite{li2023guidelines} by reporting effect sizes and confidence intervals instead. Specifically, we report group medians and Cliff's $\delta$ with 95\% confidence intervals~\cite{cliff1993dominance}; positive values indicate higher scores in the Graph group. Participant-background comparisons are descriptive.

\subsubsection{Comprehension measures and scoring}
\label{sec:study:measures:primary}

The comprehension test contained 9 questions in 5 parts worth 23 points. Parts~1--3 formed the 16-point primary outcome and assessed comprehension of the features participants asked the agent to build. Part~4 formed a 5-point secondary outcome about a feature participants had not modified, and Part~5 provided a 2-point exploratory score about execution order. Participants completed the test after Task~3 with continued read-only access to their assigned interface, as described in \autoref{sec:study:procedure}.

The test drew on Letovsky's account of programmers' knowledge of a program's purpose, implementation, and their relationship~\cite{letovsky1987}, and Sillito et al.'s~\cite{sillito2006} catalogue of questions programmers ask while modifying code. Questions used behavior-level language and did not identify files, extensions, classes, functions, or variables. As part of the test, participants rated their perceived project understanding on a 7-point scale and their confidence in each answer from 0 to 10. We treat these self-reported measures as exploratory complements to the scored comprehension outcome.

The rubric divided each question into 1-point rows awarded when an answer expressed the required idea, independent of wording. Three independent blinded runs of \texttt{claude-opus-5} scored each rubric row across all 16 de-identified responses in shuffled order. The runs agreed unanimously on 354 of 368 decisions (Fleiss' $\kappa=.943$); a fourth blinded run adjudicated the 14 non-unanimous decisions. No independent human rater scored the responses.

\subsubsection{Task performance and experiential measures}

We recorded acceptance-test results and elapsed time for each task. We defined task correctness as passing the complete acceptance-test suite within the 15-minute task period and task completion time as the elapsed time from the beginning of a task until the complete suite first passed. We summarized completion time by task and overall, with overall time calculated as the sum of each participant's three task durations. Because a technical recording issue left one Graph-group participant's Task~1 duration missing, we retained that participant's available task durations but excluded them from analyses requiring an overall completion time.

The post-task survey included 6 adapted NASA-TLX items scored from 0 to 6~\cite{hart1988tlx,hart2006tlx}. Higher values indicated greater mental demand, physical demand, temporal demand, effort, and frustration. Perceived performance ranged from 0 (perfect) to 6 (failure), so lower values indicated better perceived performance. Four 5-point TXAI items adapted from Choudhuri et al.~\cite{choudhuri2025attention} measured trust in the coding agent. We reverse-scored wariness and summed the four items into a 4--20 composite, with higher values indicating greater trust. One participant selected \emph{I'm not sure} for predictability, which we treated as missing from that item and from the complete-case composite.

\subsubsection{Interaction logs and participant accounts}

The application logged messages and file operations in both groups and Graph interactions in the Graph group. We reconstructed 24 participant--task timelines for the Graph group and organized them around requests to the agent. Two logs contained short export gaps, so event counts are lower bounds. The post-task survey contained 3 items about reliance and 8 open-ended questions adapted from the After-Action Review for AI~\cite{dodge2021aarai}.

We used an LLM-assisted hybrid deductive--inductive framework analysis to relate these records to RQ2, drawing on recent human-in-the-loop approaches that use multiple model passes alongside researcher review~\cite{choudhuri2026copilot}. The RQ1 observations served as sensitizing concepts rather than conclusions to be confirmed. A primary pass coded all 24 timelines and the 8 Graph-group response sets by oversight phase, purpose, Graph mechanism, channel transition, verification behavior, perceived control, and friction. It also retained uses and explanations outside this initial framework. A separate run independently coded a stratified set of 12 timelines spanning all tasks and levels of Graph use, and another audited event reconstruction and claim-to-source traceability across all 24 timelines. We used these checks to refine the codebook and identify counterexamples; we did not treat the runs as independent human coders or calculate intercoder agreement. Quotations were selected after this synthesis and checked against the source responses.

As a descriptive check, we examined whether the relationship-oriented comprehension and later-task pace patterns appeared on both sides of the sample median in programming experience.

\begin{table*}[t]
  \centering
  \caption{Evaluation measures and analytic roles. Higher scores indicate more of the named construct unless otherwise specified. Group-level measures compare the Graph and Chat-only groups unless marked as applying only to the Graph group.}
  \label{tab:measures}
  \begin{tabular}{@{}cl>{\raggedright\arraybackslash}p{0.48\textwidth}>{\raggedright\arraybackslash}p{0.23\textwidth}@{}}
    \toprule
    RQ & Role & Measure & Range or unit \\
    \midrule
    RQ1 & \primaryrole & Task-related comprehension & 0--16 points \\
    RQ1 & \secondaryrole & Unseen-feature comprehension & 0--5 points \\
    RQ1 & \exploratoryrole & Execution-order comprehension & 0--2 points \\
    RQ1 & \exploratoryrole & Relationship and exact-location rubric rows & 0--18 and 0--5 points \\
    RQ1 & \exploratoryrole & Confidence and self-rated understanding & 0--10 and 1--7 \\
    RQ1 & \secondaryrole & Task correctness & Complete acceptance-test suite passed within 15 minutes \\
    RQ1 & \secondaryrole & Task duration & Minutes; lower is faster \\
    RQ1 & \secondaryrole & Adapted NASA-TLX items & 0--6; higher is worse \\
    RQ1 & \secondaryrole & Trust items and planned composite & 1--5; composite 4--20 \\
    \midrule
    RQ2 & \exploratoryrole & Graph actions and action sequences (Graph group only) & Counts, order, and phase \\
    RQ2 & \exploratoryrole & Reliance items and AAR-AI participant accounts & 5-point items and open-ended responses \\
    \bottomrule
  \end{tabular}
  \Description{A four-column table mapping each measure to a research question and an analytic role. The columns are the research question, the role, the measure, and its range or unit. Eleven measures are listed. One is primary, five are secondary and five are exploratory. The primary measure is task-related comprehension, scored out of 16. Three secondary measures concern the tasks. They are unseen-feature comprehension out of 5, task correctness, and task duration in minutes where lower is faster. Task correctness records whether the complete acceptance-test suite passed within fifteen minutes. The other two secondary measures are self-reported. The adapted workload items are scored 0 to 6, where a higher value is worse. The trust items are scored 1 to 5, with a composite from 4 to 20. Three exploratory measures come from the comprehension test. They are execution-order comprehension out of 2, and the relationship and exact-location rubric rows out of 18 and 5. The third is confidence and self-rated understanding, on 0 to 10 and 1 to 7 scales. The last two exploratory measures are counts and sequences of actions on the Graph, and the reliance items together with open-ended participant accounts. The action measures apply to the Graph group alone.}\end{table*}

\subsection{Procedure}
\label{sec:study:procedure}
\label{sec:study:ethics}

We conducted 2 pilot sessions to verify the procedure and session length, then clarified the session overview, interface guide, and survey questions. Pilot data were excluded from the evaluation.

Sessions ran remotely on participants' computers and were screen-recorded with consent. After a 15-minute setup period and the warm-up, participants completed the 3 study tasks for 15 minutes each. They then exported the application log and task results to the moderator before completing the comprehension test for approximately 20 minutes. During the test, participants could navigate and inspect content already available, but they could not send new instructions to the agent or modify the project. They then completed a 10-minute survey about workload, trust in the coding agent, and their experience overseeing agent-generated changes. The moderator provided technical help and kept time but did not suggest strategies, evaluate actions, or request think-aloud narration.

\paragraph{Ethics.} Our institution's institutional review board approved the study. Participants provided verbal consent and separate consent for screen recording with audio. We stored data under participant identifiers, removed identifiers and group labels from comprehension responses before scoring, and retained study data according to the approved protocol. Participants received \$40 in gift cards and/or Claude Pro subscription credit.

%

\section{Results}
\label{sec:results}

All 16 participants completed the study. Overall, the Graph group showed stronger relationship-oriented comprehension and a steadier task pace, alongside higher mental demand, lower effort, and lower trust. Participants used the Graph mainly to inspect project structure and changes, while relying on chat to direct the agent and other tools for verification. We next describe these findings in more detail.

\subsection{RQ1: Project comprehension, task performance, cognitive load, and trust}
\label{sec:results:rq1}

The RQ1 results reveal differences in what participants understood, how their pace developed across tasks, and how they experienced the work. Although the groups had similar primary project-comprehension scores and everyone completed all three tasks correctly, Graph participants demonstrated more relationship-oriented comprehension and a steadier pace across later tasks, alongside higher mental demand, lower effort, similar frustration, and less trust in the coding agent (\autoref{tab:rq1} and \autoref{tab:rq2}).

\begin{table*}[t]
\centering
\small
\caption{Comprehension and task duration by group. Medians are based on eight participants per group. $\delta$ is Cliff's delta, positive when the Graph group had the higher value. The relationship and exact-location rows partition the complete 23-point rubric and therefore overlap with the task-related, unseen-feature, and execution-order outcomes; the rows are not additive. A positive later-task change means that the later tasks took longer.}
\label{tab:rq1}
\begin{tabular}{@{}p{0.43\textwidth}lccc@{}}
\toprule
Outcome & Role & Graph & Chat-only & $\delta$ [95\% CI] \\
\midrule
Task-related comprehension, 0--16 & \primaryrole & 6.0 & 5.5 & $+0.03\;[-0.49,+0.53]$ \\
Unseen-feature comprehension, 0--5 & \secondaryrole & 2.5 & 0.5 & $+0.39\;[-0.23,+0.78]$ \\
Execution-order comprehension, 0--2 & \exploratoryrole & 0.5 & 0.0 & $+0.22\;[-0.28,+0.63]$ \\
Finding relationships among components, 0--18 & \exploratoryrole & 7.5 & 4.0 & $+0.45\;[-0.15,+0.81]$ \\
Finding exact code locations, 0--5 & \exploratoryrole & 0.5 & 2.0 & $-0.45\;[-0.80,+0.12]$ \\
\midrule
\multicolumn{5}{@{}l}{\textit{Descriptive measures}} \\
Confidence per item, 0--10 & \exploratoryrole & 5.4 & 4.9 & $+0.30\;[-0.28,+0.71]$ \\
Self-rated understanding, 1--7 & \exploratoryrole & 3.6 & 3.4 & $+0.16\;[-0.38,+0.62]$ \\
\midrule
\multicolumn{5}{@{}l}{\textit{Task duration, minutes; lower is faster}} \\
Task~1 & \secondaryrole & 10.1 & 9.5 & $+0.14\;[-0.44,+0.64]$ \\
Task~2 & \secondaryrole & 10.3 & 14.5 & $-0.53\;[-0.86,+0.13]$ \\
Task~3 & \secondaryrole & 9.6 & 11.9 & $-0.50\;[-0.83,+0.08]$ \\
Later-task change, mean of Tasks~2 and~3 minus Task~1 & \exploratoryrole & 0.4 & 4.5 & $-0.80\;[-0.95,-0.37]$ \\
\bottomrule
\end{tabular}
\Description{A five-column table of twelve outcomes in three blocks, comprehension, descriptive measures, and task duration. The columns are the outcome with its range, its analytic role, the Graph group median, the Chat-only group median, and Cliff's delta with a ninety-five percent confidence interval. Only one row is primary, task-related comprehension, where the two groups are close at 6.0 against 5.5. The Graph group is higher on unseen-feature comprehension at 2.5 against 0.5, on execution order at 0.5 against zero, and on finding relationships at 7.5 against 4.0. It is also higher on confidence at 5.4 against 4.9 and on self-rated understanding at 3.6 against 3.4. The Chat-only group is higher on finding exact code locations at 2.0 against 0.5. In the task-duration block the Graph group is slower on Task 1 at 10.1 minutes against 9.5. It is faster on Task 2 at 10.3 against 14.5 and on Task 3 at 9.6 against 11.9. The later-task change is 0.4 minutes for the Graph group and 4.5 for the Chat-only group. Every confidence interval in the table includes zero except the one on that last row, which runs from minus 0.95 to minus 0.37.}
\end{table*}

\paragraph{Primary comprehension outcome.} On the 16-point score covering the features participants built, the median was 6.0 in the Graph group and 5.5 in the Chat-only group ($\delta = +0.03$, 95\% CI $[-0.49,+0.53]$; \autoref{tab:rq1}). This estimate provides little evidence of a difference in overall task-related comprehension. However, the aggregate score combines questions about component relationships and exact code locations, potentially obscuring differences in what participants understood.

\paragraph{Relationship-oriented comprehension.} We therefore examined how each group earned its task-related points. Although both groups earned 44 points in total, 37 concerned relationships among components in the Graph group, compared with 27 in the Chat-only group; exact-location points showed the reverse pattern (7 versus 17). The participant-level comparisons across the complete rubric followed the same direction for relationships ($\delta=+0.45$, 95\% CI $[-0.15,+0.81]$) and exact locations ($\delta=-0.45$, 95\% CI $[-0.80,+0.12]$; \autoref{tab:rq1}). Thus, the observed distinction concerns what participants understood---relationships rather than locations---not greater overall comprehension.

\paragraph{Task performance.} All 48 participant-task attempts passed their acceptance-test suites within the time limit. Due to the recording issue described in \autoref{sec:study:measures}, one Graph-group participant's Task~1 elapsed time was missing. The Graph group had a higher observed median duration on Task~1 (10.1 versus 9.5 minutes) and lower observed medians on Tasks~2 and~3 (10.3 versus 14.5 minutes and 9.6 versus 11.9 minutes, respectively; \autoref{tab:rq1}). Among participants with complete timing data, every participant in the Chat-only group took longer on the later tasks than on Task~1, with a median increase of 4.5 minutes, whereas changes in the Graph group clustered near zero. Because the tasks were cumulative and both groups completed them in the same order, the contrast reflects how their pace changed as the project evolved: Graph participants maintained their pace, whereas Chat-only participants slowed. This pattern suggests that the Graph helped participants carry project understanding from earlier tasks into later ones rather than reconstructing it for each task.

\begin{table*}[t]
\centering
\small
\setlength{\tabcolsep}{6pt}
\caption{Self-reported workload and trust by group. $\delta$ is Cliff's delta, positive when the Graph group had the higher value. Workload items run 0 to 6 and a higher value is the worse outcome, with performance running from Perfect at 0 to Failure at 6. Trust items run 1 to 5 and a higher value is more trust, and the composite runs 4 to 20. Medians are over eight participants per group, except predictability and the composite, where one Chat-only participant answered \emph{I'm not sure} and seven remain.}
\label{tab:rq2}
\begin{tabular}{@{}lcccccc|c@{}}
\toprule
& \multicolumn{6}{c|}{\textbf{Cognitive load (adapted NASA-TLX)}} & \textbf{Trust} \\
\cmidrule(lr){2-7} \cmidrule(l){8-8}
Measure & Mental & Physical & Temporal & Perform. & Effort & Frustr. & Composite \\
\midrule
Cliff's delta ($\delta$) & $+0.14$ & $-0.44$ & $-0.06$ & $+0.20$ & $-0.38$ & $0.00$ & $-0.55$ \\
\midrule
\multicolumn{8}{@{}l}{\textit{Median values for each group}} \\
Graph & 2.0 & 1.0 & 1.0 & 3.5 & 1.0 & 1.0 & 17.0 \\
Chat-only & 1.5 & 1.0 & 1.0 & 1.0 & 1.5 & 1.0 & 18.0 \\
\bottomrule
\end{tabular}
\Description{The table is transposed, with seven measures as columns and three rows of values beneath them. The first six columns are the workload dimensions, mental demand, physical demand, temporal demand, performance, effort, and frustration. The seventh, set apart by a rule, is the four-item trust composite. Reading across in that order, Cliff's delta is plus 0.14, minus 0.44, minus 0.06, plus 0.20, minus 0.38, exactly zero, and minus 0.55. The Graph medians are 2.0, 1.0, 1.0, 3.5, 1.0, 1.0, and 17.0. The Chat-only medians are 1.5, 1.0, 1.0, 1.0, 1.5, 1.0, and 18.0. The two groups share the same median on three of the six workload dimensions. Apart from perceived performance in the Graph group, every workload median is 2.0 or lower.}
\end{table*}

\paragraph{Cognitive load.} The Graph group reported slightly higher mental demand ($\delta=+0.14$), lower effort ($\delta=-0.38$), and similar frustration ($\delta=0.00$; \autoref{tab:rq2}). This pattern suggests greater cognitive engagement without a corresponding increase in effort or frustration. In RQ2, we examine how participants' accounts of structural engagement and interface unfamiliarity may explain this combination.

\paragraph{Trust.} The planned complete-case trust composite was lower in the Graph group (median 17 versus 18; $\delta=-0.55$; \autoref{tab:rq2}). The largest observed item-level difference was confidence in the assistant, with medians of 4 in the Graph group and 5 in the Chat-only group ($\delta=-0.50$, 95\% CI $[-0.80,+0.00]$); the other item estimates were smaller and uncertain. One Chat-only participant was excluded from the composite after selecting \emph{I'm not sure} for predictability. The RQ2 accounts identify several possible contributors, but they do not establish that lower trust means participants were thinking more.

Taken together, the RQ1 findings suggest that \toolname{} changed how participants engaged with the project. Graph participants developed more relationship-oriented comprehension and maintained their pace across cumulative tasks; at the same time, they reported higher mental demand. Compared with the Chat-only group, their lower effort and similar frustration suggest that this greater mental demand did not make the interaction more effortful or frustrating, while their lower trust shows that greater visibility into the project and the agent's activity was not accompanied by greater confidence in the system. This combination suggests that a shared graph representation can support project comprehension and sustained task performance while keeping developers cognitively involved in oversight. Descriptively, the relationship-oriented comprehension and sustained-pace patterns appeared in both the lower- and higher-experience subgroups. We return to this design trade-off in the Discussion.


\subsection{RQ2: Oversight during creation and review}
\label{sec:results:rq2}
\label{sec:results:graphuse}

The RQ2 findings show a division of labor between the Graph and other parts of the interface. Participants used the Graph primarily to inspect project structure and follow where the agent was working, while chat remained their main channel for directing the agent. The Graph supported orientation and awareness of where changes occurred, but participants still relied on chat, code, tests, and diffs to understand or verify implementation details.

\paragraph{Inspecting project structure and agent activity.} Of 241 focal Graph actions, 205 were module clicks or detail openings. Across 23 participant--task segments with Graph activity, the first Graph action occurred before an instruction in 7, while the agent was working in 12, and after completion in 3; the remaining segment began with a Graph-initiated instruction. P4 described the Graph as making it ``easier to see the overall structure of the code and where changes were made. Rather than clicking on a file for context, you can click on a module.'' Live activity cues also supported monitoring: P3 felt most in control when the interface highlighted the component being worked on, and P5 valued the animated marker showing which component was changing. Yet this location awareness did not always explain the agent's process. P3 lost track during long autonomous sequences, P1 could not see the agent's reasoning, and P5 could not inspect the Graph while it reorganized after a change.

\paragraph{Directing changes through chat.} Of 47 requests to the agent, 43 originated in chat. The remaining four were Graph-initiated edits, all from P4; P2 once dragged a module into chat as context, and P5 initiated but canceled a Graph edit. P4 likewise reported feeling most in control in chat, while P8 explained, ``When the change affects several stages in the diagram, I am not sure which step should start first. Pasting the task directly into the chat box makes it easier.'' These records position the Graph as an inspection and contextualization surface rather than a replacement for natural-language instructions.

\paragraph{Reviewing changes across tools.} Of the 47 requests, 26 edited at least one file and 19 produced a Graph update naming affected modules. Before the next request, participants directly clicked or opened an affected module after 6 updates and investigated an affected module or relationship through search after 3 more. Six participants opened the change tracker 12 times, and five selected tracker entries 22 times. Even so, participants located correctness checks outside the Graph: P4 wanted to write tests the agent could not see, P6 wanted diffs, and P2 would inspect complex components before relying on the agent. No Graph participant would allow these changes to a personal project unconditionally: four said no, and four limited acceptance to lower-stakes cases or prior checks. At the same time, all eight acknowledged accepting at least some study changes without really checking them. The Graph provided resources for review, but prewritten tasks and visible passing tests still allowed participants to delegate without verifying every change.

\paragraph{Carrying project context forward.} Five participants used the Graph before their first instruction in at least one later task, accounting for 7 of 16 Task~2 and Task~3 segments. In Task~3, P1 searched across previously introduced TOC and Panel components before examining Structure; P4 replayed earlier Extension System changes before initiating the next change from that module; and P2 carried a previously inspected module into chat as context. These actions show how participants could use the Graph to carry project understanding forward rather than reconstruct it for each task.

Together, the logs and participant accounts help explain the RQ1 patterns. The Graph maintained a project-level view that participants revisited as the project evolved, directing attention toward component relationships and supporting continuity across tasks. At the same time, interpreting this representation required attention, while missing reasoning, diffs, tests, and unclear Graph provenance limited participants' confidence in the system. The Graph therefore supported oversight by maintaining project context and showing where the agent was working, while chat and conventional development resources remained necessary for directing and verifying its work.

\section{Discussion}
\label{sec:discussion}
Our results suggest that \toolname{} changed how participants engaged with an evolving project. Although the groups had similar primary project-comprehension scores, Graph participants showed more relationship-oriented comprehension and maintained their pace across cumulative tasks. They used the Graph mainly to inspect project structure and agent-generated changes, while using chat to direct the agent and conventional development resources to verify its work. We discuss what this combination suggests for other software-engineering activities, the design of shared representations, and human oversight of coding agents.

\subsection{From Project Comprehension to the Software-Engineering Lifecycle}

Although we evaluated \toolname{} during project comprehension and modification, the same need for structural orientation occurs elsewhere in the software-engineering lifecycle. Code review is one promising example. Understanding a submitted change is a central challenge in review, particularly when reviewers are unfamiliar with the affected files or must determine whether a change breaks code elsewhere~\cite{bacchelli2013codereview,tao2012understandchanges,pascarella2018infoneeds}. This problem may become more pronounced when agents generate more code while leaving developers with less understanding of how it fits together~\cite{afroz2026fastspurious}.

A live Graph could complement a diff by carrying the structural context of a change. A diff shows which lines changed; the Graph shows where those lines belong in the system and what the affected modules connect to. A reviewer could therefore use the Graph to orient themselves before examining implementation details in the diff. However, the Graph currently shows where a change occurred, not why it was made. Because reviewers also need the rationale for a change~\cite{bacchelli2013codereview,ram2018reviewability}, an interface for review would need to connect the Graph with the agent's plan, conversation, or change description. We did not evaluate code review, so this is a possible application of the design rather than a finding of the present study.

\subsection{Design Implications for Shared Representations}

Our results offer initial evidence about the four design requirements embodied in \toolname{}: a structural, live, manipulable, and adaptive representation. Because the study compared the complete Graph interface with the Chat-only interface, it cannot isolate the contribution of any individual requirement or feature.

A structural representation may support orientation across files and components~\cite{storey1999cognitive,brooks1983theoryofcomprehensionofcomputerprograms,sillito2006}. Although the groups had similar overall task-related comprehension scores, the Graph group earned more points about relationships among components, whereas the Chat-only group earned more about exact code locations. This distinction matters because a forgotten location can often be recovered through search, while a missing account of how several components fit together may require reading across files. The observed pattern identifies problem-level structure as a promising target for comprehension support.

A live representation may help developers carry project context forward as an agent modifies the code~\cite{horowitz2023liverichcomposablequalities,xie2024waitGPT}. The Graph remained synchronized with the project and marked modules affected by the agent's work. Participants inspected affected modules after updates, described the activity markings as useful for following the agent, and sometimes revisited the Graph before beginning a later task. These behaviors help explain how Graph participants maintained their pace as the cumulative tasks progressed.

A manipulable and adaptive representation may complement natural-language interaction~\cite{shneiderman1983directmanipulation,shu2023sensecape,lee2025HINT}. Participants primarily clicked modules and opened details; they used search, Focus mode, dragging, and Graph-initiated editing much less often. Natural-language chat remained easier for expressing changes involving several stages, while the Graph helped participants identify modules and provide project context. The results therefore support treating language and direct manipulation as complementary channels rather than replacing chat with a visual interface~\cite{cohen1992role,masson2024directgpt}. They provide less evidence about the value of adaptive detail or Graph-based editing, which would require more targeted evaluation.

\subsection{Human Factors in Agentic Coding Oversight}

Oversight requires developers to understand enough of the project to determine whether intervention is warranted. Yet agentic workflows make this harder by distributing activity across planning, execution, tool use, and generated changes~\cite{passi2025oversight,mitchell2026aiagentspush,dhanorkar2026humanoversight}. Our results suggest that supporting this understanding does not simply mean reducing cognitive work. Graph participants reported higher mental demand but lower effort and similar frustration, suggesting that interpreting the representation kept them cognitively involved without making the interaction feel more effortful or frustrating. The Graph supported this involvement through a persistent view of where the agent was working and how that work related to the rest of the project.

Greater visibility was not accompanied by greater trust. Participants could see where the agent was working, but the Graph did not always explain its reasoning or provide the diffs and tests they wanted for verification. Participants therefore returned to chat or code when the Graph lacked sufficient detail and described diffs and independent tests as necessary for checking the agent's work. No participant in the Graph group was unconditionally willing to let the agent modify their own project without checking. The Graph supported orientation and monitoring while preserving a role for independent verification.

These findings also offer a design perspective on cognitive debt. Storey~\cite{storey2026technicaldebtcognitiveintent} describes cognitive debt as system understanding eroding faster than it is replenished when developers accept code they did not construct themselves. The steadier task pace and participants' reuse of the Graph are consistent with carrying structural understanding forward as the project changed. A live structural representation may support this continuity without requiring developers to stop the agent after every generated change. This approach complements practices that ask developers to perform additional comprehension work, such as walkthroughs or explanation gates~\cite{storey2026technicaldebtcognitiveintent,sankaranarayanan2026mitigatingepistemicdebtgenerative}. Because our short study did not measure cognitive debt over time, longitudinal work is needed to determine whether this support preserves project comprehension across sessions.

\section{Limitations and Future Work}

\subsection{Scope of the Prototype}

\toolname{} is a research prototype that instantiates the four design requirements examined in this paper. We built it to carry this study rather than to release it, and we evaluated it with a single project and coding agent. We have not run it on codebases of other sizes, other languages, or other structures. We therefore make no claim about how far it scales, what it would cost to run, or whether the same design would hold for a different kind of project.
Two implementation constraints leave those questions open. The whole uploaded project is sent to the model in one request, so a project has to fit inside that request, and the cost of drawing the Graph grows with the size of the project. We also do not store the Graph between sessions. Within a session the Graph is held steady. After an edit we send the current version back with the request and ask for the modules that did not change to be returned as they were. Opening the same project again may be different, because the Graph is written from scratch, so the wording of the module names and captions may not be the same. We have not measured how much the decomposition itself varies. Besides, we also hold the overview to eight modules so it stays readable at a glance, which means a large project is compressed harder than a small one.

\subsection{Limitations of the Study}
\paragraph{Task Design.}
Participants in both groups described pasting the task brief straight into the agent. If a task can be finished without the developer forming an account of how the system works, then neither interface is being asked to support that account, and a difference between them has little room to appear. Our tasks were small, self-contained, and came with acceptance tests, so a participant could pass one to the agent in a single message without first working out which parts of the system it touched. Writing tasks that resist being delegated in one step is the first thing we would change, and we think it is a general difficulty in evaluating tools of this kind.

\paragraph{Sample and Measurement Constraints.}
Our sample was small. As stated in \autoref{sec:study:analysis}, the resulting estimates have wide confidence intervals and may be unstable; similar observed values do not establish that the groups are equivalent. The comprehension test measured performance with continued read-only access to the assigned interface rather than unaided retention. Participants also used \toolname{} after only a brief warm-up, so unfamiliarity may have influenced the workload and trust ratings. Two interaction logs contained short export gaps, making the reported Graph-event counts lower bounds.
Task correctness told us less than we expected. Every task attempt passed its acceptance tests, so this measure had no variation and could not separate the groups.
We examined many measures, and the difference in how much participants slowed after Task~1 was not specified in advance. We therefore treat it as exploratory and as the clearest candidate for targeted replication in a later study. Random assignment also left the groups somewhat unbalanced on some background variables, as can occur in a sample this small.

\paragraph{Evaluation of a Bundled Design.}
The design space for a live Graph that a coding agent acts on has not been mapped, and we did not set out to map it. We built one point in it and did not compare our choices against alternatives, so we cannot say which of them matter and which could have been made differently. Participants primarily navigated modules and opened their details, while using search, Focus mode, dragging, and Graph-initiated editing less often. We therefore evaluated the interface as a bundle and cannot determine which design requirement or interaction produced an observed pattern.

\subsection{Future Work}
Our findings point toward shared representations as persistent project memory. Future systems could preserve the Graph across sessions and connect project structure with the rationale, provenance, tests, and unresolved decisions behind each change. Such a representation could support not only agentic implementation, but also code review, debugging, maintenance, onboarding, and handoffs among developers or agents.

A second direction is to develop the Graph from an inspection surface into a medium for oversight and coordination. Developers could use it to express constraints, compare proposed plans, inspect evidence for changes, and correct the agent's understanding of the project. Adaptive views could present different levels of detail for different tasks or roles while preserving a stable underlying representation shared by everyone involved.

Evaluating this vision requires longitudinal and team-based studies that examine whether shared representations help developers regain context, recognize when intervention is needed, verify changes, and transfer understanding to others. These studies should use larger and more varied projects, tasks requiring cross-component reasoning, and component-level comparisons that isolate the contributions of the four design requirements.

\section{Conclusion}
We presented \toolname{}, an agentic coding interface built around a structural, live, manipulable, and adaptive graph representation. In our 16-participant evaluation, the groups had similar primary project-comprehension scores, but Graph participants demonstrated more relationship-oriented comprehension and maintained their pace across cumulative tasks. They used the Graph mainly to inspect project structure and changes, while using chat to direct the agent. Their higher mental demand and lower trust, alongside lower effort and similar frustration, suggest continued cognitive involvement rather than increased confidence in automation. Shared representations could similarly support code review, debugging, maintenance, and other activities involving agent-generated work. More broadly, effective oversight of agents requires not only records of their activity, but persistent representations of how artifacts under human responsibility are evolving, so people can understand, question, and redirect changes as they unfold.

\bibliographystyle{ACM-Reference-Format}
\bibliography{reference}
\end{document}